\def\year{2021}\relax
\documentclass[letterpaper]{article} 
\usepackage{aaai21}  
\usepackage{times}  
\usepackage{helvet} 
\usepackage{courier}  
\usepackage[hyphens]{url}  
\usepackage{graphicx} 
\usepackage{subcaption}
\usepackage{amsmath}
\usepackage{amsfonts}
\usepackage{lipsum}
\usepackage{natbib}  
\usepackage{caption} 
\DeclareMathOperator*{\argmax}{arg\,max}
\DeclareMathOperator*{\argmin}{arg\,min}

\title{From Behavioral Theories to Econometrics:\\Inferring Preferences of Human Agents from Data on Repeated Interactions}
\author {
    Gali Noti \\
}

\affiliations{
    Harvard University\\ 
		The Hebrew University of Jerusalem \\
    galinoti@gmail.com
}

\begin{document}

\maketitle

\begin{abstract}
We consider the problem of estimating preferences of human agents from data of strategic systems where the agents repeatedly interact. Recently, it was demonstrated that a new estimation method called ``quantal regret'' produces more accurate estimates for human agents than the classic approach that assumes that agents are rational and reach a Nash equilibrium; however, this method has not been compared to methods that take into account behavioral aspects of human play. In this paper we leverage equilibrium concepts from behavioral economics for this purpose and ask how well they perform compared to the quantal regret and Nash equilibrium methods. We develop four estimation methods based on established behavioral equilibrium models to infer the utilities of human agents from observed data of normal-form games. The equilibrium models we study are quantal-response equilibrium, action-sampling equilibrium, payoff-sampling equilibrium, and impulse-balance equilibrium. We show that in some of these concepts the inference is achieved analytically via closed formulas, while in the others the inference is achieved only algorithmically. We use experimental data of 2x2 games to evaluate the estimation success of these behavioral equilibrium methods. The results show that the estimates they produce are more accurate than the estimates of the Nash equilibrium. The comparison with the quantal-regret method shows that the behavioral methods have better hit rates, but the quantal-regret method performs better in terms of the overall mean squared error, and we discuss the differences between the methods.
\end{abstract}

\section{Introduction} \label{sec:intro}

Suppose that we are looking at data generated by several human agents repeatedly playing a game. 
In the data we can observe the actions played by each player in each period of play. 
However, the data do not contain the players' true preferences (``values''), which are the private information of each player. 
In many multi-agent strategic applications 
there are large amounts of such data (e.g., bidding data from online auctions), and 
these private values of the agents, which actually drive the actions they choose to play, 
are required as a crucial first step before performing any analysis on the data. 
For example, the values are needed to determine whether the game results are efficient, 
whether the game rules can be improved, or in order to test any 
theoretic behavioral model on the data.

In order to 
draw estimates of the values 
from the data, 
it is necessary to make assumptions about the way the players behave. 
The classic approach to this econometric estimation task 
is to assume that the players are in a Nash equilibrium. This assumption provides mathematical equations that allow us to extract estimates for the parameters of interest. 
However, behavioral studies have demonstrated how  
Nash equilibrium 
does not always 
describe human play (see, e.g., \cite{handbook,goeree2001}). 

In a recent study, a new estimation method called ``quantal regret'' was suggested as more suitable for the case of human players \cite{quantalregret}.
While rational players are expected to minimize their regret in the repeated game, the quantal-regret method 
only assumes that players are more likely to act in a way that gives them lower regret.\footnote{The regret is in the usual sense used in the regret-minimization literature \cite{BM2007}. This notion assumes that players manage to achieve at least as much utility as they could have gotten from playing any {\em fixed} action repeatedly.
}
The authors demonstrated on data from controlled experiments with human players---on the 2x2 game dataset of \cite{Selten2008} and on the auction dataset of \cite{www2014}---that the quantal-regret method provides significantly more accurate estimates than both the classic Nash equilibrium-based methods and the method that assumes perfect regret minimization \cite{eva}. 
The advantage of the quantal-regret method over the regret-minimization method  
was further confirmed on ad auction field data in \cite{noti2020}. 
See Appendix \ref{ne} for an overview of the Nash equilibrium method and Appendix \ref{qr} for an overview of the quantal-regret method. 

In this work we study the benefits from using models of behavioral equilibria for the purpose of the estimation task. 
Similar to the classic approach of Nash equilibrium, the approach of using the behavioral equilibria assumes that the players are at a stationary state; however, these stationary states are defined in terms of behavioral considerations that have been studied in the behavioral literature and shown to describe human biases. 
Thus, these behavioral equilibria have the potential to be suitable models for deriving estimates from data of human players.  
We develop new econometric estimation methods that are based on behavioral equilibrium models, and ask:  
How well do these behavioral estimation methods succeed in providing accurate estimates of the preferences of human players?

\begin{figure}[t]
\begin{subfigure}{.5\linewidth} 
  \includegraphics[scale=0.57]{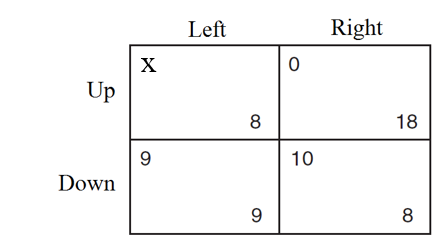} 
  \caption{}
  \label{fig:example-mtx-x-y}
\end{subfigure}%
\begin{subfigure}{.5\linewidth} 
  \includegraphics[scale=0.57]{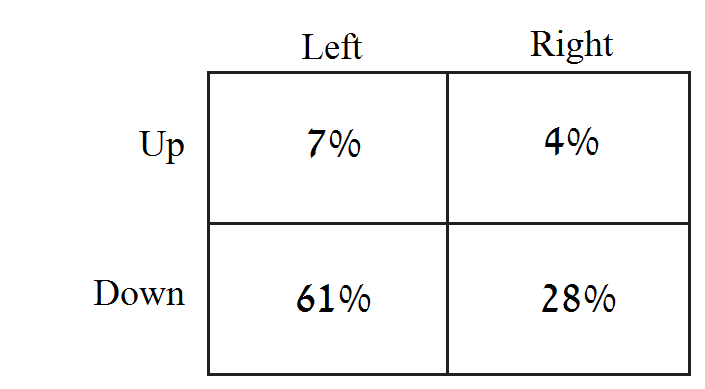}
  \caption{}
  \label{fig:example-freq}
\end{subfigure}
\caption{An example of the estimation task for a 2x2 game. (\ref{fig:example-mtx-x-y}) The utility matrix of Game 1 in Selten and Chmura (2008), with the parameter that we wish to estimate hidden by an $x$. 
The upper-left and lower-right corners in each cell are the utilities of the row and column players, respectively. 
(\ref{fig:example-freq}) The average empirical frequency that was obtained in one of the sessions by human agents playing Game 1 repeatedly for 200 periods. 
}
\label{fig:example}
\end{figure}

The concepts that we study are the four behavioral equilibria that were analyzed in \cite{Selten2008}: 
(1) quantal-response equilibrium (QRE), which extends Nash equilibrium by allowing the players to make mistakes \cite{qre}; 
(2) payoff-sampling equilibrium (PSE), in which players optimize against samples of their payoffs from each of their own pure strategies \cite{pse}; 
(3) action-sampling equilibrium (ASE), in which players optimize against samples of strategies played by the other players \cite{Selten2008}; 
and (4) impulse-balance equilibrium (IBE), in which players respond to impulses of the difference between the payoff that could have been obtained and the payoff that has been received, such that for all players expected upward impulses are equal to expected downward impulses \cite{ibe}. 

\cite{Selten2008} compared the ability 
of these models to {\em predict} human behavior in repeated 2x2 games with unique mixed strategy equilibria. 
They tested the different models on data from experiments of 2x2 games that they ran with human participants. 
Two-player 2x2 games are a simple strategic setting and serve as a natural testing ground for a comparison of alternative models as they have been shown to give rise to various behavioral biases \cite{camerer2011}.
The comparison of \cite{Selten2008}
showed that all four behavioral concepts predicted human play better than the Nash equilibrium model. 

We develop four econometric {\em estimation} methods based on each of the four behavioral equilibrium models. 
The estimation task is defined for general normal-form games, similar to the task considered in \cite{quantalregret} for 2x2 games: given the empirical frequencies of players' actions when repeatedly playing a normal-form game, and given the utilities defining the game except for a single unknown utility of one of the players, we need to estimate the remaining unknown utility. 
Figure \ref{fig:example} shows an example of the estimation task for a 2x2 game. 
We show that for the impulse-balance equilibrium and the quantal-response equilibrium the estimation can be performed analytically via closed formulas. 
By contrast, for the two sampling models---the payoff-sampling equilibrium and the action-sampling equilibrium---there is no closed expression for the estimated utility. Thus, we suggest an algorithmic approach to performing the estimation based on these two models.

We test the success of the behavioral estimation methods, in comparison with the 
quantal-regret 
and Nash equilibrium methods, 
on the 2x2 game dataset from 
\cite{Selten2008}. 
Specifically, we ``hide from ourselves'' the utilities in the game, one at a time, estimate the hidden utility from the observed data, and then compare the estimate to the true value that was actually used in the experiment.\footnote{Notice that such a comparison requires the use of experimental data; it cannot be performed with real-world data, where the true values of the players 
are unknown.}
The results show that the behavioral equilibrium methods manage to produce better estimates than the Nash equilibrium model that assumes rationality of the players. 
The comparison with the quantal-regret method shows differences in the estimation-error patterns between the methods: on the one hand, the four behavioral methods manage to have a higher frequency of close hits than the quantal-regret method, but, on the other hand, they also have a non-negligible number of large errors that decrease their overall performance, leading to a better overall performance of quantal regret in terms of mean squared error. These results suggest that combining the methods, perhaps by incorporating behavioral biases into the quantal-regret method, can improve the estimation results.

The contribution of the paper is twofold.
The first contribution is conceptual: the basic idea of using the four behavioral equilibrium models to devise new methods to answer econometric questions. While the behavioral literature analyzes and discusses these behavioral equilibrium concepts, the novelty in the present paper is that they can be used in the {\em inverse} direction for the econometric estimation purpose. 
This paper develops four new econometric methods for estimating missing utilities in normal-form games, based on four established behavioral equilibrium models, and  
exemplifies how to use these methods in 
2x2 games. 
It shows that behavioral theories can indeed be useful for econometrics---definitely beating the classic rationality-based econometric method (the Nash estimates), and also have an advantage 
over 
the quantal-regret method in terms of hit rates. 
The second contribution is the comparison between different modeling concepts for the econometric estimation task when the players are humans. Quantal regret is a new model and method, and 
although it has been compared to rational models, it has not yet been compared to any {\em behavioral} models. 
For a method that claims to be suitable for econometrics with human players, the comparison with methods that are based on leading behavioral models is called for, as they are the ``real competition.'' In this paper we perform such a comparison for the first time, and confirm, based on the same dataset and settings that were used to demonstrate the quantal-regret method itself in \cite{quantalregret}, that quantal regret indeed works well for econometric estimation with human players, even when compared to the behavioral models.

The rest of the paper is organized as follows.  
In the next section we formally specify our setting and overview additional related literature. 
In Section \ref{sec:methods} we describe in more detail each of the behavioral equilibrium concepts that we study and show how to derive its corresponding estimation method. 
In Section \ref{sec:results} we evaluate the estimation success of the methods on data from experiments with human players, and conclude in Section \ref{sec:conclusions}.

\section{Preliminaries}

\subsection{Setting} \label{setting} 

For the general estimation setting, consider $n$ agents repeatedly playing a normal-form game. 
Denote by $A_i$ the action space of player $i$ and by $a^t_i \in A_i$ the action played by player $i$ at period $t$. 
For all players $i\in\{1,...,n\}$, we denote by $m_i=|A_i|$ the number of available actions. 
The unknown information that we wish to estimate is captured by a parameter $\theta \in {\Theta \subseteq \mathbb{R}^d}$, 
and the game is defined by known utility functions $u_i(a_i,\vec{a}_{-i},\theta)$, where $\vec{a}_{-i}$ denotes an action profile of all the players except for player $i$. 
We are given the empirical sequence of actions that the players actually played in $T$ repetitions of the game, i.e., $\tilde{a} = ((a^1_1,...,a^1_n),(a^2_1,...,a^2_n),...,(a^T_1,...,a^T_n))$. Our goal is to estimate the unknown parameter $\theta$ from the observed behavior $\tilde{a}$, the observed utility functions, and any prior $p(\theta)$ we may have on the possible values of $\theta$. 

We will evaluate the estimation methods in the 2x2-game setting, where the utility function is defined by a fixed 2x2 utility matrix, and there are two players---the row player and the column player---who are repeatedly playing according to this function (see example in Figure \ref{fig:example-mtx-x-y}). 
In each game period, each player has a binary choice: the row player chooses an action in $A_{r}=\{Up, Down\}$ and the column player chooses an action in $A_{c}=\{Left, Right\}$. We will also use the abbreviations $U, D, L, R$ to denote these pure strategies (actions). 
The unknown information, which we wish to estimate, is simply one of the eight parameters in the utility matrix, 
e.g., the utility $x$ of the row player from playing $U$ when the column player plays $L$ 
(as in Figure \ref{fig:example-mtx-x-y}).

We denote by $\vec{p}_i$ the mixed strategy of player $i$ (i.e., $\vec{p}_i$ is a probability distribution over $A_i$) in which she plays action 
$j\in A_i$ with probability $p_{ij}$. 
Denote by $u_i(a,\vec{p}_{-i})$ the expected utility of player $i$ from playing action $a$ when the other players play the mixed strategies $\vec{p}_{-i}=(\vec{p}_1, ...,\vec{p}_{i-1}, \vec{p}_{i+1},...,\vec{p}_n)$.
For 2x2 games, we denote by $p_U$ the mixed strategy of the row player in which she plays $U$ with probability $p_U$, and by $p_L$ the mixed strategy of the column player in which she plays $L$ with probability $p_L$.

\subsection{Estimation Framework} \label{framework} 

We define the econometric estimation task for normal-form games, similar to the definition considered by \cite{quantalregret} for 2x2 games for presenting the quantal-regret method.
Denote by $x$ the utility of player 1 from the action profile $(a_{11},a_{21},...,a_{n1})$ in which each of the $n$ players plays her first action. For the special case of a 2x2 game, $x$ will denote the utility of the row player from the action profile $(U,L)$, as in Figure \ref{fig:example-mtx-x-y}. 
We describe each method for the case of estimating 
$x$ (the generalization to the estimation of any of the other utilities in the game is straightforward),
such that each of the 
estimation methods can be fitted in the following framework:

\noindent\fbox{%
    \parbox{0.45\textwidth}{%
		\vspace{3px}
		\textbf{The Framework for the Estimation Methods: }\\
        \textbf{Input}: 
				\begin{enumerate}
				\item The empirical frequencies $\tilde{p}_1, ..., \tilde{p}_n$ in which each player $i$ played each of her possible actions (i.e., $\tilde{p}_i = (\tilde{p}_{i1},...,\tilde{p}_{im_i})$). 
				\item The utility matrix of the game except for the single missing utility $x = u_1(a_{11},a_{21},...,a_{n1})$. 
				\end{enumerate} 
\textbf{Output}: an estimate $\hat{x}$ of the missing utility $x$. 
   }%
}


\subsection{Additional Related Literature}

The importance of behavioral modeling has been well recognized in the AI literature in various contexts that involve humans, e.g., in the context of security games with bounded rational adversaries \cite{nguyen2013,yang2013,kar2015}, automated navigation amongst human agents \cite{ziebart2009,bera2017,fisac2018}, 
	modeling advertiser behavior in online advertising auctions \cite{rong2016,noti2020}, 
	and human computation systems \cite{kamar2012,mao2013,yin2015}.

Settings of strategic interactions between human players are widely studied in behavioral economics, and the subfield of behavioral game theory addresses the gap between the standard game-theoretic modeling of rational utility-maximizing agents and actual human behavior; 
	for a broad introduction to this literature, see \cite{camerer2011,handbook} and references therein. 
Works in behavioral economics \cite{Selten2008,erev1998,camerer1999} and, more recently, in the AI literature \cite{wright2010,wright2019} have studied the ability of behavioral models to predict actions of human players in games. In addition, a recent line of work has studied the ability of machine-learning methods to predict human play in games \cite{hartford2016,kolumbus2019,fudenberg2019,noti2016,plonsky2017}.
In the prediction task considered in these works, all the game parameters are known and are used to predict players' actions. 
By contrast, in the econometric estimation task that we consider the behavioral models are used in the inverse direction, such that the actions played by the players are known and are used to estimate unknown parameters of the game (and in particular the players' utilities in the game).

The task of learning game parameters from observed data has primarily been studied under the standard game-theoretic assumption that the players are in a Nash equilibrium, e.g., in \cite{vorobeychik2007,honorio2015,athey2010,varian2007}. 
\cite{eva} proposed to estimate advertisers' values from their bidding data in sponsored search auctions by relying on a weaker rationality assumption that the players minimize their regret over time. 
\cite{www2017} experimentally evaluated this ``min-regret'' method with human players and showed
that the regret-minimization assumption was sufficient to produce value estimates that are at least as accurate as the estimates of classic equilibrium-based methods; 
however, 
the improvement was not significant, and players' actions were only partially consistent with the regret-minimization assumption.  

Relaxations of the rationality assumption, for the purpose of learning game parameters from observed data, were considered in \cite{quantalregret} and \cite{ling2018}. 
The QRE relaxation of the Nash equilibrium model proposed by \cite{qre} was used in \cite{ling2018} to learn game parameters in zero-sum non-repeated games; however, their method was not evaluated on real behavioral data and therefore it 
remained unclear whether the QRE modeling is useful for 
estimating parameters with human players. 
In \cite{quantalregret}, the quantal-regret method was proposed as a general method that extends the min-regret method of \cite{eva} by relaxing the regret-minimization assumption, and was shown to significantly improve the estimation performance on experimental data with human players in comparison with the min-regret and Nash equilibrium-based methods.  However, the quantal-regret method has not been compared with 
any behavioral method that takes into account behavioral aspects of human play. 
See Appendix \ref{qr} for an overview of the quantal-regret method. 
As explained in Section \ref{sec:intro}, in this paper we leverage equilibrium concepts from behavioral economics for the purpose of estimating game parameters from players' observed actions in repeated games, and evaluate their performance in comparison with the quantal-regret and Nash
equilibrium methods on data from controlled experiments with human players.

\section{The Behavioral Estimation Methods} \label{sec:methods} 

In this section we present the econometric methods that we develop based on the behavioral equilibrium concepts,
according to the estimation framework described in Section \ref{framework}.

\subsection{Quantal-Response Equilibrium} \label{sec:qre}

\textbf{The Logic: }
In quantal-response modeling players choose with higher probabilities actions that give them higher expected payoffs. 
In the common exponential form of this modeling it is assumed that players are exponentially more likely to play actions with higher expected payoffs.
The model has a parameter $\lambda$, which multiplies the expected payoff in the exponent. 
This parameter determines the rationality of the player:  for $\lambda \rightarrow \infty$ players become completely rational and choose the action that maximizes their payoffs, while for $\lambda$ close to zero the payoffs do not affect their choice and they choose an action uniformly at random. 

In a quantal-response equilibrium (QRE) \cite{qre}, each player (correctly) takes the mistakes of the others into account and quantally best responds to the behavior of the others.  
Therefore, in a quantal-response equilibrium, the following equation holds simultaneously for all players $i$ and actions $j\in A_i$:

\begin{equation} \label{eq:qre}
p_{ij} = \frac{e^{\lambda u_i(j,\vec{p}_{-i})}}{\sum_{j^{'}\in A_i} e^{\lambda u_i(j^{'},\vec{p}_{-i})}}
\end{equation} 

For example, for 2x2 games, where there are two players and two actions for every player, the two following equations hold simultaneously in a quantal-response equilibrium:

\begin{subequations}
\begin{equation} \label{eq:qre1}
p_U = \frac{e^{\lambda u_{r}(U,p_L)}}{e^{\lambda u_{r}(U,p_L)} + e^{\lambda u_{r}(D,p_L)}}
\end{equation}
\begin{equation} \label{eq:qre2}
p_L = \frac{e^{\lambda u_{c}(L,p_U)}}{e^{\lambda u_{c}(L,p_U)} + e^{\lambda u_{c}(R,p_U)}}
\end{equation}
\end{subequations}

For the 2x2 game experimental data of \cite{Selten2008} $\lambda=1.05$ gave the best fit, i.e., minimized the sum of mean squared distances from the empirical choice frequencies over all games investigated in the experiment.\footnote{In \cite{Selten2008} some of the model fits were incorrect. In our analyses we use the corrected fits that were also pointed out by \cite{comment}.} 

\noindent \textbf{The QRE Estimation Method: }
If we assume that the players were in a quantal-response equilibrium, then Equation \ref{eq:qre} allows us to extract an estimate for $x$ analytically, by substituting $i=1$ and $j=1$, as well as all the other terms that are given in our estimation framework: the empirical frequencies $\tilde{p}_{-i}$ of the other players, the other utilities of player 1, 
and the model parameter $\lambda$.
For example, for the case of 2x2 games, Equation \ref{eq:qre1} 
depends on $x$ (since the expected utility is $u_r(U,p_L) = p_L \cdot x + (1-p_L) \cdot u_r(U,R)$)), and we can extract an estimate for $x$ by substituting all the other terms as follows:
\begin{equation*}
\resizebox{0.47\textwidth}{!}{$\hat{x} = \frac{ \lambda \cdot \left(u_r(D, L)\cdot \tilde{p}_L + u_r(D, R) \cdot (1-\tilde{p}_L)\right) - \lambda \cdot u_r(U, R) \cdot (1-\tilde{p}_L) + ln(\frac{\tilde{p}_U}{1-\tilde{p}_U}) } {\lambda \tilde{p}_L} $} 
\end{equation*}

Similarly, the second equation (Equation \ref{eq:qre2}) can be used to estimate each of the utilities of the column player.


\subsection{Action-Sampling Equilibrium} \label{sec:ase}

\textbf{The Logic: }
According to the action-sampling modeling, each player takes a sample of 
observations of her opponents' past 
actions and then best responds to this sample. If there is more than a single best response, the player mixes with equal probabilities all actions that are best responses to her sample. An action-sampling equilibrium (ASE) 
describes a stationary state of 
large populations, where each player takes a sample and optimizes against it. 
The sample size $n_s$ is a parameter of this model. In the 2x2 game experiments of \cite{Selten2008} the best fit was obtained for $n_s=12$. 

We derive the action-sampling equilibrium equations for general normal-form games, similar to the equilibrium derived by \cite{Selten2008} for 2x2 games. 
Denote by $s_i(\vec{a}_{-i})$ the number of times player $i$ observed the action tuple $\vec{a}_{-i}$ of the other $n-1$ players in her sample, and  
denote by $\alpha_{ij}(s_i)$ the probability in which player $i$ chooses action $j\in A_i$ after observing 
the given sample  $s_i$. 
Since each player best responds to her sample, we know that if 
$j \in M = \argmax_{a\in A_i} \sum_{\vec{a}_{-i}\in A_{-i}} s_i(\vec{a}_{-i}) u_i(a,\vec{a}_{-i})$ then $\alpha_{ij}(s_i) = |M|^{-1}$, or otherwise $\alpha_{ij}(s_i) = 0$.
Then, the following equation describes the total choice probabilities (a set of 
$m_i-1$ equations for each of the $n$ players) 
in which player $i$ chooses action $j\in A_i$, and should hold simultaneously in equilibrium:

\begin{equation} \label{eq:ase_general}
p_{ij} = E_{s_i} \big[ \alpha_{ij}(s_i) \big]
\end{equation}

For example, for 2x2 games 
\cite{Selten2008} derive the equations for $\alpha_{ij}$ as follows:

\begin{equation*}
\resizebox{0.07\textwidth}{!}{$\alpha_{rU}(s_r) = $}
	\begin{cases}
	\resizebox{0.008\textwidth}{!}{$1$} & \resizebox{0.35\textwidth}{!}{$\text{if } s_r(L) \cdot \big(u_r(U,L) - u_r(D,L) \big) > (n_s-s_r(L)) \cdot \big(u_r(D,R) - u_r(U,R)\big)$} \\ 
	\resizebox{0.02\textwidth}{!}{$0.5$} & \resizebox{0.35\textwidth}{!}{$\text{if } s_r(L) \cdot \big(u_r(U,L) - u_r(D,L) \big) = (n_s-s_r(L)) \cdot \big(u_r(D,R) - u_r(U,R)\big)$} \\
	\resizebox{0.008\textwidth}{!}{$0$} & \resizebox{0.06\textwidth}{!}{$\text{otherwise}$}  \\
	\end{cases}
\end{equation*}

\begin{equation*}
\resizebox{0.07\textwidth}{!}{$\alpha_{cL}(s_c) = $}
	\begin{cases}
	\resizebox{0.008\textwidth}{!}{$1$} & \resizebox{0.35\textwidth}{!}{$\text{if } (n_s-s_c(D)) \cdot \big(u_c(U,L) - u_c(U,R) \big) > s_c(D) \cdot \big(u_c(D,R) - u_c(D,L)\big)$} \\ 
	\resizebox{0.02\textwidth}{!}{$0.5$} & \resizebox{0.35\textwidth}{!}{$\text{if } (n_s-s_c(D)) \cdot \big(u_c(U,L) - u_c(U,R) \big) = s_c(D) \cdot \big(u_c(D,R) - u_c(D,L)\big)$} \\
	\resizebox{0.008\textwidth}{!}{$0$} & \resizebox{0.06\textwidth}{!}{$\text{otherwise}$}  \\
	\end{cases}
\end{equation*}

Then, for the 2x2 game, 
the expectation in Equation \ref{eq:ase_general} takes the form of a binomial distribution 
as follows:  

\begin{subequations}
\begin{equation} \label{eq:ase1}
p_U = \sum_{k=0}^{n_s} \binom{n_s}{k} p_L^k(1-p_L)^{n_s-k}\alpha_{rU}(s_r)
\end{equation}
\begin{equation} \label{eq:ase2}
p_L = \sum_{l=0}^{n_s} \binom{n_s}{l} p_U^l(1-p_U)^{n_s-l}\alpha_{cL}(s_c)
\end{equation}
\end{subequations}
where $k = s_r(L)$ and $l = s_c(D)$.

\noindent \textbf{The ASE Estimation Method: }
We follow our estimation framework (described in Section \ref{framework}) and estimate the utility $x=u_1(a_{11},a_{21},...,a_{n1})$ of player 1. 
From the 
set of 
equations that Equation \ref{eq:ase_general} specifies, the $m_i-1$ equations for which $i=1$ depend on $x$, and we use them for the estimation. 
Note that when the empirical distribution of play is an exact equilibrium, all these equations yield the same estimate for $x$, but 
when the empirical play is not an exact equilibrium there may be different estimates from different equations, 
and we set the final estimate for $x$ as the average of these estimates. 
In this case, the standard deviation of the estimates 
indicates how far the empirical distribution was from an ASE and 
may be used for error estimation. 

In contrast to the methods in Sections \ref{sec:qre} and \ref{sec:ibe}, our sampling equations 
do not depend continuously on $x$, and 
we cannot 
derive a closed expression for $x$. 
Therefore, we take an algorithmic approach to estimate $x$. 
Specifically, $x$ is estimated as the value for which the $m_i-1$ equations of Equation \ref{eq:ase_general} with $i=1$ are satisfied with the minimal error, by carrying the following procedure: 

\noindent For $j \in \{1,...,m_1-1\}$, compute an estimate $\hat{x}_j$ as follows:
\begin{enumerate}
	\item Compute $\hat{p}_{1j}(x)$ as the right-hand side of Equation \ref{eq:ase_general} for a grid of possible values 
	of $x$, by substituting the available empirical frequencies and the known utilities of player 1 (see the input description in Section \ref{framework}).
	\item Choose as the $j$th estimate $\hat{x}_j$ the value of $x$ that minimizes the distance from the empirical frequency, i.e., $\hat{x}_j = \argmin_x |\hat{p}_{1j}(x) - \tilde{p}_{1j}|$. If the minimum is obtained for a range of values choose the middle point. 
\end{enumerate}
Finally, set as the estimate $\hat{x}$ the average over the $m_1-1$ estimates $\hat{x}_j$.

Note that this algorithmic approach is similar to the approach used for the quantal-regret estimate calculation \cite{quantalregret}, where the regret was computed for the different values of the estimated parameter. See Appendix \ref{qr} for more details on the quantal-regret method.

For example, for a 2x2 game, we use Equation \ref{eq:ase1} to estimate the utility $x=u_r(U,L)$ of the row player algorithmically, as follows: 
compute $\hat{p}_U(x)$ as the right-hand side of Equation \ref{eq:ase1} for different possible values of $x$ (and substitute the available empirical frequency $\tilde{p}_L$ and the three known utilities of the row player), and finally choose $\hat{x} = \argmin_x |\hat{p}_U(x) - \tilde{p}_U|$.


\subsection{Payoff-Sampling Equilibrium}

\textbf{The Logic: }
This concept is based on the assumption that each player takes samples of equal size, one for each of her own available actions, 
and then plays the action of the sample with the highest payoff sum (or chooses at random one of the maximizing actions if there is more than one) \cite{pse}.
Similar to the sampling equilibrium concept described in Section \ref{sec:ase}, the payoff-sampling equilibrium (PSE) describes a stationary state of large populations in which each player optimizes against her samples. 
The sample size $n_s$ is a parameter of this model. The value $n_s=6$ for each of the samples gave the best fit for the experimental 2x2 game data of \cite{Selten2008}.


We can derive the equations for the payoff-sampling equilibrium for general normal-form games, in a similar analysis to that described in Section \ref{sec:ase} for the ASE, and similar to the PSE derived by \cite{Selten2008} for 2x2 games.
Specifically, let $s_i(\vec{a}_{-i}, j)$ be the number of times player $i$ observed the action tuple $\vec{a}_{-i}$ of the other $n-1$ players in her sample for her action $j \in A_i$. 
Similar to Section \ref{sec:ase}, 
we can specify the probability $\alpha_{ij}(s_i)$ in which player $i$ chooses action $j \in A_i$ for the sample $s_i$ of each of her actions, 
based on the assumption that each player plays the action with the highest payoff sum. 
Then, the total probability in which player $i$ chooses action $j$ is the expectation over the samples; this should hold simultaneously for all players $i$ and actions $j \in A_i$ in equilibrium.

For example, for 2x2 games, the payoff-sampling equilibrium for two specific players in two populations that play the 2x2 game are as follows:
let $k_U = s_r(L,U)$ and $k_D = s_r(L,D)$ be the number of times the row player observed $L$ in her sample for her actions $U$ and $D$, respectively, and 
let $l_L = s_c(U,L)$ and $l_R = s_c(U,R)$ be the number of times the column player observed $U$ in her sample for her actions $L$ and $R$, respectively.
Therefore, the following equations describe the total choice probabilities of the two players, and should hold simultaneously in equilibrium: 

\begin{subequations}
\begin{equation} \label{eq:pse1}
\resizebox{.99\hsize}{!}{$p_U = \sum_{k_U,k_D=0}^{n_s} \binom{n_s}{k_U} \binom{n_s}{k_D} p_L^{k_U+k_D}(1-p_L)^{2n_s-k_U-k_D} \alpha_{rU}(s_r)$}
\end{equation}
\begin{equation} \label{eq:pse2}
\resizebox{.99\hsize}{!}{$p_L = \sum_{l_L,l_R=0}^{n_s} \binom{n_s}{l_L} \binom{n_s}{l_R} p_U^{l_L+l_R}(1-p_U)^{2n_s-l_L-l_R}\alpha_{cL}(s_c)$}
\end{equation}
\end{subequations}

\noindent \textbf{The PSE Estimation Method: }
The estimation of $x=u_1(a_{11},a_{21},...,a_{n1})$ (according to our estimation framework) is performed similarly to the estimation method for the action-sampling equilibrium described in Section \ref{sec:ase}, since also here the equilibrium equations that depend on $x$ do not allow us to extract $x$ analytically and instead require an algorithmic approach. 
Specifically, we estimate $x$ as the value for which the $m_1-1$ equilibrium equations for player 1 are satisfied with the minimal error, by carrying out the procedure described in Section \ref{sec:ase}: 
for $j \in \{1,...,m_1-1\}$ compute $\hat{x}_j$ from the equilibrium equation for $p_{1j}$ for different possible values of $x$, and by choosing $\hat{x}_j = \argmin_x |\hat{p}_{1j}(x) - \tilde{p}_{1j}|$, and set the final estimate for $x$ as the average over $\hat{x}_j$. 
For example, for the 2x2 game, estimate $x=u_r(U,L)$ by computing $\hat{p}_U(x)$ as the right-hand side of Equation \ref{eq:pse1} for different possible values of $x$, and then choose as the estimate $\hat{x} = \argmin_x |\hat{p}_U(x) - \tilde{p}_U|$.


\subsection{Impulse-Balance Equilibrium} \label{sec:ibe}

\textbf{The Logic: }
The impulse-balance equilibrium (IBE) was proposed by \cite{ibe} 
and is a ``semi-quantitative'' version of the learning direction theory of 
\cite{SB1999}.  
According to the learning direction theory, when a decision maker repeatedly makes choices on the same parameter and receives feedback, she gets impulses: if a higher (lower) parameter would have brought a higher payoff she gets an upward (downward) impulse. The theory assumes that the decision maker will tend to choose in the direction of the impulse. 
An impulse-balance equilibrium is a stationary distribution, in which the expected upward impulses are equal to the expected downward impulses for each of the players simultaneously. To reflect loss aversion as in prospect theory \cite{prospecttheory}, in this model losses are counted twice in the computation of impulses: a loss is counted once as a part of the foregone payoff and once more due to being a loss.

The IBE model is suitable for settings in which 
a decision maker repeatedly makes choices on the same parameter, 
such as bidding in auctions or binary choice problems. 
Therefore, 
we will focus on normal-form games with $m=2$ actions for each player, 
where 
the probability of playing one of the pure strategies is the parameter the players choose 
(i.e., $p_{i1}$ for each player $i$, and in two-player 2x2 games: $p_U$ and $p_L$ for the row and column players, respectively). 
As suggested by \cite{Selten2008}, the utility of the pure strategy maximin---which is the ``security level'' a player can ensure by playing a pure action 
no matter what the other players do---may naturally serve as the reference level for determining losses. 

\begin{figure}[t]
	\centering
		\includegraphics[scale=0.36]{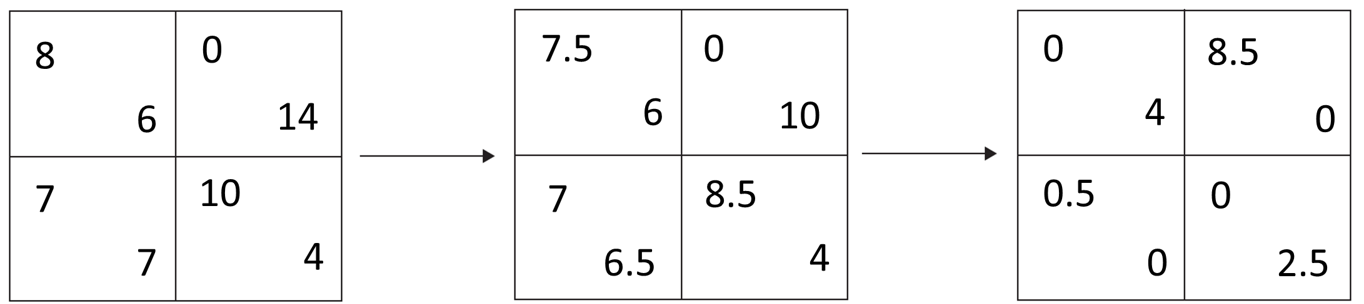}
	\caption{Construction of the transformed and impulse matrices: (1) the original game; (2) the transformed form of the game; (3) the impulse matrix of the game (based on an example from Selten and Chmura (2008)).}
	\label{fig:transformed}
\end{figure}

\noindent \textbf{The IBE Estimation Method: }
The estimation of the utility $x=u_1(a_{11},a_{21},...,a_{n1})$ of player 1 (according to our estimation framework described above) is performed by deriving the equilibrium equations for player 1, which in turn requires computing the impulses from one action to another. 
We compute the impulses by using the ``transformed form'' of the game as suggested by \cite{Selten2008}.   
Specifically, the estimation process involves four steps: 
\begin{enumerate}
	\item Construct the transformed game: Denote by $s_i$ the security level of player $i$. 
The transformed game matrix is constructed from the utility matrix of the game such that 
for each player $i$ and actions $a_i \in A_i$ and $\vec{a}_{-i} \in A_{-i}$: 
\begin{equation*}
\resizebox{.96\hsize}{!}{$u^{tr}_i(a_i,\vec{a}_{-i}) = 
	\begin{cases}
	u_i(a_i,\vec{a}_{-i}) &\text{if $u_i(a_i,\vec{a}_{-i}) \leq s_i$}  \\
	\frac{1}{2} \big[s_i + u_i(a_i,\vec{a}_{-i})\big] &\text{otherwise}  \\
	\end{cases}$}
\end{equation*}
Figure \ref{fig:transformed} exemplifies the construction for Game 3 from the experiment of \cite{Selten2008}. 
	
	\item Construct the impulse matrix: Player $i$ receives an impulse in the direction of her other action if and only if after a play she observes that she could have obtained a higher utility by choosing her other action. The size of the impulse is the foregone utility in the transformed game (which thus gives the losses a double weight).  
	That is, the impulse matrix is computed such that for each player $i$ and actions $a_i \in A_i$ and $\vec{a}_{-i} \in A_{-i}$: 
\begin{equation*}
\resizebox{.96\hsize}{!}{$imp_i(a_i,\vec{a}_{-i}) = 
	\begin{cases}
	\Delta=u^{tr}_i(A_i\setminus\{a_i\},\vec{a}_{-i})-\\u^{tr}_i(a_i,\vec{a}_{-i}) &\text{if $\Delta > 0$}  \\
	0 &\text{otherwise}  \\
	\end{cases}$}
\end{equation*}
See the example in the second step of Figure \ref{fig:transformed}. 

	\item Derive the impulse-balance equilibrium equations: 
	The equilibrium conditions require that for each of the players the expected impulses will be balanced. 
	For example, in the 2x2 game, for the row player in an impulse-balance equilibrium the expected impulse from $Up$ to $Down$ is equal to her expected impulse from $Down$ to $Up$. 
	That is, 
	\begin{equation*}
	\begin{split}
		p_L\cdot imp_r(U,L) + (1-p_L) \cdot imp_r(U,R) = \\
		p_L\cdot imp_r(D,L) + (1-p_L) \cdot imp_r(D,R)
	\end{split}
	\end{equation*}

	\item Extract an estimate for $x$ by substituting the empirical frequencies and the known utilities of player 1.

\end{enumerate}

Note that the exact formula for $x$ depends on the structure of the game. 
The reason is that the transformed and impulse matrices depend on the determination of the security level $s_i$. 
Thus, in order not to assume any prior knowledge of the structure of the game, the implementation of the estimation process requires us to consider all the different cases that result in different security levels. 

\section{Experimental Evaluation} \label{sec:results}

\begin{table*}[t]
\centering

\resizebox{0.6\textwidth}{!}{ 
\begin{tabular}{|l|l|l|l|l|l|l|}
\hline
\multicolumn{7}{|c|}{}\\[-0.8em]
\multicolumn{7}{|c|}{2x2 Games -- Over All 108 Sessions} \\  \hline
& & & & & & \\[-0.8em]
	& QR & ASE & PSE & IBE & QRE & NE \\ \hline
	& & & & & & \\[-0.7em]
RMSE  & 2.32 & 2.39 & 2.59 & 2.81 & 2.82 & 3.41 \\
Average Error  & 2.09 & 2.04 & 2.18 & 2.44 & 2.25 & 2.99 \\
Std Error  & 1.58 & 1.72 & 1.97 & 2.20 & 2.20 & 2.32 \\ 
$\pm 3$ Hit Rate  & 81.13\% & 85.42\% & 84.95\% & 82.06\% & 82.52\% & 68.87\% \\
\hline
\end{tabular}
}
\caption{Evaluation results of the six estimation methods over all 108 sessions of the 2x2 game dataset. \label{tbl:results}}
\end{table*}

In this section we 
evaluate the estimation success of the four behavioral econometric methods proposed in Section \ref{sec:methods} 
in comparison with the quantal-regret (QR) method and with the method that is based on Nash equilibrium (NE). 

We perform the evaluation by using the data from the experiment of \cite{Selten2008}; the same dataset 
was used both 
in \cite{Selten2008} to evaluate the {\em prediction} success of the behavioral equilibria models and 
in \cite{quantalregret} to evaluate the {\em estimation} success of the quantal-regret method. 
In the experiment, 12 2x2 games were investigated. 
Each of the games
had non-negative utilities
and was ``completely mixed,'' i.e., 
had a unique equilibrium in which each player
plays each of her actions with positive probability, according to all equilibrium concepts discussed in the present paper.
There were 12 independent subject groups (``sessions'') for each
of games 1--6, and 6 
for each of
games 7--12, for a total of 108 sessions. In each session there were 4 row players and
4 column players, who repeatedly played a game over 200
periods. 
See \cite{Selten2008} for more details on
the experimental setup and 
for the utility matrices of the games.
The data consist of the utility matrices of the 12 games, as well as of the empirical frequencies of play 
for each player over her 200 plays. 
See illustration in Figure \ref{fig:example}. 

We applied each of the six methods to estimate the 8 parameters defining the game (4 utilities of the row player and 4 of the column player), one at a time, and compared the estimates with the parameters that were actually used in the experiment. 
As in \cite{Selten2008} and \cite{quantalregret}, we consider the ``session level'' of the experimental data. 
That is, for each independent session we estimate the 8 parameters 
by considering the average of the empirical frequencies of play of the 8 players in the session. 

We define the estimation error of an estimate $\hat{x}$ for a utility of which the true value is $x$ as $error(\hat{x}) = |\hat{x} - x|$. 
Our main measure of success is the root mean squared error (RMSE)
achieved by each method. 
Specifically, the RMSE of a set of estimates $S$ is $RMSE(S) = \sqrt{\frac{1}{|S|} \sum_{\hat{x} \in S} error^2(\hat{x})}$.
We also measure the ``$\pm 3$ hit-rate,'' which is the percentage of estimates within a distance of $3$ from the true value. Note that in our estimation context exact hits are hard to achieve, and thus some interval from the true value is taken. We choose an interval of 3, which is about 15\% of the utility range of the games in our dataset, for the sake of comparison with the results of \cite{quantalregret} who used this specific measure. To complete the picture, we also compare the full distributions of the estimation errors of the methods. 

In the implementation of the four behavioral methods we use the same parameters as in \cite{Selten2008,comment}. 
The values were $\lambda=1.05$ for the QRE method, and 
the sampling sizes were 
$n_s=6$ and $n_s=12$ for the PSE and ASE methods, respectively. 
The implementation of the NE and the QR methods was according to the procedure described in \cite{quantalregret}. 
For the QR method, this includes a regret-aversion parameter $\lambda=3$ (according to their suggested rule of thumb), and the same uniform prior over the parameter range $[0,22]$ (in intervals of 0.01).
We use the same prior also for the two proposed algorithmic methods ASE and PSE, and for a fair comparison we restrict the estimates of the other methods to the same parameter range.\footnote{The values $0$ and $22$ were the minimal and maximal utilities (respectively) in all the games investigated in the experiment. We also tried wider ranges of up to 40, which increased the errors for all methods as could be expected. However, this did not change our conclusions, except for the IBE method which was more sensitive to the utilized prior and had a larger error than the NE method.} 

Table \ref{tbl:results} presents the estimation results over all 108 sessions in the experiment for the six estimation methods.
It can be seen that in terms of RMSE the QR method performs better than the four behavioral methods, which in turn outperform the NE method. 
Specifically, the RMSE of each of the four behavioral methods is significantly lower than the RMSE of the NE method (paired two-sided Wilcoxon signed rank test, $N=108$ sessions, $p < 0.002$), and the RMSE using QR is lower than that of each of the four behavioral methods, but the difference is statistically significant only in comparison with the IBE method ($p < 0.02$). The gap between QR and the behavioral methods is large except for ASE 
whose RMSE is close to the QR method. 

\begin{figure*}[t]
	\centering
		\includegraphics[scale=0.45]{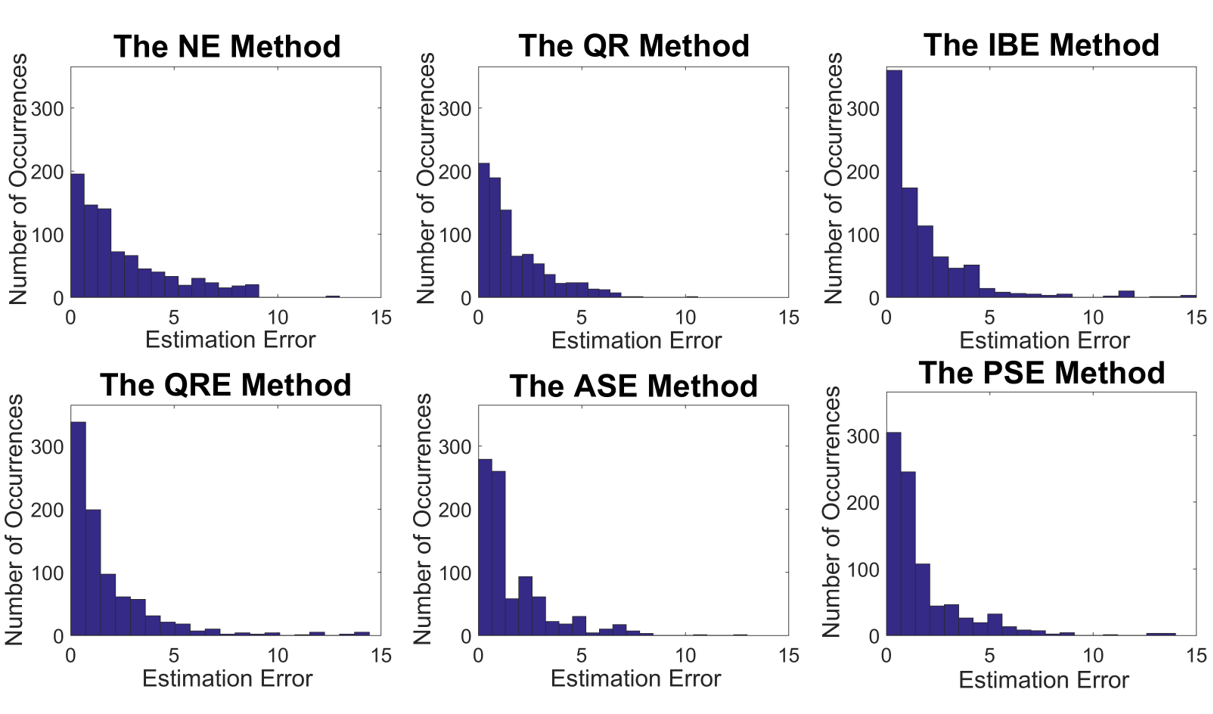}  
	\caption{Error distributions: Histograms of the estimation errors for each of the six estimation methods.}
	\label{fig:dists}
\end{figure*}

The comparison of the four behavioral methods shows that the two sampling methods perform better than the IBE and QRE methods. 
Over all 108 sessions, the RMSE of each of the sampling equilibrium methods is significantly lower than the RMSE of the IBE method (paired two-sided Wilcoxon signed rank test, $N=108$ sessions, $p < 0.002$), but the difference from the QRE method is not statistically significant.
Note that this 
is different from the comparison results of \cite{Selten2008} 
who found that in the context of the prediction task the four behavioral equilibrium models underlying our econometric methods were not significantly different. It is possible that the econometric task studied in the present paper is more sensitive than the prediction task to the differences between these models and allows a better discrimination between them.

So far we have seen that the QR method, which uses the rule-of-thumb regret-aversion parameter $\lambda=3$, has better RMSE 
than the methods that are based on behavioral equilibria. 
We also tried another variant of the QR method in which $\lambda$ was directly learned from the data in cross-validation. 
Specifically, for each of the 12 games we learned $\lambda$ by a direct fit of the empirical regret results obtained on the other 11 games. 
This gave 12 values of $\lambda$ with an average of $2.64$ and a small standard deviation of 0.12. However, using QR with this learned parameter for each game did not change the results, and the QR method still outperformed the four behavioral methods in terms of the RMSE. 

Let us now take a closer look 
at 
the distributions of the estimation errors obtained by the six methods. 
Table \ref{tbl:results} presents the average and standard deviations of the distribution for each of the methods, as well as their $\pm 3$ hit rates. 
Figure \ref{fig:dists} plots the full distribution of the estimation errors. 
It can be seen in the plots that the four behavioral methods very frequently hit close to the target, but on the other hand have a non-negligible frequency of large errors. 
By contrast, the QR method less often hits the target exactly, but usually does not go too far and does not make large errors. 
These shapes of the distributions explain both the advantage of the QR method in terms of RMSE and the standard deviation of the errors, and the disadvantage in terms of hit rates, which are presented in Table \ref{tbl:results}. 

\section{Conclusions} \label{sec:conclusions}

We have shown 
that it is possible to leverage equilibrium concepts from behavioral economics to infer the preferences of human players in normal-form games, 
and have shown how to derive the econometric estimation methods from four established behavioral equilibrium models. 
When the equilibrium conditions did not provide a direct analytic solution, we showed how to perform the estimation by taking an algorithmic approach. 
In fact, the two behavioral algorithmic methods that we proposed performed better than the analytic ones. 
All four behavioral estimation methods that we studied 
outperformed the Nash equilibrium method that assumes that the players are rational. This is consistent with previous literature that suggests that these models 
better capture human behavior. 

The comparison with the quantal-regret method highlights both the advantages of this method and the potential for improvements based on behavioral considerations. 
First, the success of the quantal-regret method shows that quantal regret provides a suitable modeling for human strategic behavior for the estimation task  
not only compared to rational models but also compared to well-studied behavioral models. 
	Furthermore, unlike the behavioral methods, the quantal-regret method does not require any specific analysis to the estimation setting and works exactly the same in other settings such as auctions. 
Second, we observed some ``tradeoff'' between exact hits and low overall error: 
	the quantal-regret method 
	does not make large errors, but on the other hand has fewer exact hits. 
	By contrast, the behavioral methods, which make more specific assumptions about the behavior of the players at stationary states, hit very frequently close to the target, but on the other hand also make large errors.  

We believe that further research that will apply well-established knowledge from behavioral disciplines has the potential 
to improve the econometric analysis of game data that are generated by human players. 
While we focused on four behavioral equilibrium models 
it is interesting to study the estimation performance of additional behavioral models -- both equilibrium models and dynamic learning models. It is also interesting to extend the methods to more complex game settings, where multiple equilibria may exist, such as estimating players' private values from auction data, or to settings with more than a single unknown parameter, which will require searching a larger space of values. 
Such research can lead to 
an improved modeling that will break the tradeoff described above, i.e., a modeling that will have the high hit rates of the behavioral methods while still avoiding large errors like the quantal-regret method.

\section*{Acknowledgments}
This project has received funding from the European Research Council (ERC) under the European Union's
Horizon 2020 research and innovation programme (grant agreement No 740282). 
The author would like to thank Thorsten Chmura for sharing the 2x2
game dataset, Judith Avrahami and Yaakov Kareev for inspiring suggestions, 
and the four anonymous referees for helpful comments.

\appendix

\section*{Appendices}

\section{The Nash Equilibrium Estimation Method} \label{ne} 
Nash equilibrium is a standard concept in game theory to analyze game outcomes. 
It is a stationary state in which each player perfectly best responds to the other players. 
That is, Nash equilibrium is a strategy profile $(\vec{p}_1,\vec{p}_2,...,\vec{p}_n)$, in which $\vec{p}_i$ maximizes the expected utility for each player $i$, given the other players' strategies $\vec{p}_{-i}$. 
Therefore, Nash equilibrium provides a set of linear equations that allow to extract an estimate for $x$ by substituting the empirical frequencies and the known utilities which are given in our estimation framework that is described in Section \ref{framework}. 

For example, for the 2x2 game setting, a Nash equilibrium is a strategy profile $(p_U,p_L)$, in which $p_U$ and $p_L$ maximize the expected utility of the row and the column players, respectively, given the other player's strategy, and 
the following two equations (one for each of the two players) should hold simultaneously:

\begin{subequations}
\begin{equation} \label{eq:ne1}
p_L = \frac{u_r(D,R) - u_r(U,R)}{x - u_r(D,L) + u_r(D,R) - u_r(U,R)}
\end{equation}
\begin{equation} \label{eq:ne2}
p_U = \frac{u_c(R,D) - u_c(L,D)}{u_c(L,U) - u_c(R,U) + u_c(R,D) - u_c(L,D)}
\end{equation}
\end{subequations}

Thus, if we assume that the players are in a Nash equilibrium, 
$x$ can be estimated from equation \ref{eq:ne1} by substituting the other terms (assuming that $\tilde{p}_L > 0$), which are given in the estimation framework. 

\section{The Quantal Regret Estimation Method} \label{qr} 
The quantal regret modeling was suggested by \cite{quantalregret}. Its basic assumption is that players 
choose with (exponentially) higher probabilities actions that give them lower regret. This is instead of assuming perfect regret minimization, which is expected from rational players. 
The regret of a player in a repeated game is defined as the difference between the utility she could have obtained had she played the best fixed strategy in hindsight and the utility she actually obtained in the repeated game \cite{BM2007}.

In our present estimation setting (as described in Section \ref{framework}), the quantal regret method for estimating $x = u_1(a_{11},a_{21},...,a_{n1})$ first computes the regret of player 1 for each possible value of $x$,\footnote{The calculation is performed for some grid of values of $x$, in some valuation range. See \cite{quantalregret} for 
more details.}
\begin{equation*}
\resizebox{0.48\textwidth}{!}{$regret_1(x,\tilde{a}) = \frac{1}{T}\left(max_{a'_1 \in A_1} \sum_{t=1}^T u_1(a'_1,\tilde{a}_{-1}^t,x)
 - \sum_{t=1}^T u_1(\tilde{a}_1^t,\tilde{a}_{-1}^t,x)\right)$}
\end{equation*}
Then, it sets the estimate as the weighted average of the possible values of $x$, with weights that are exponentially decreasing with the regret. Specifically, assuming a prior of a uniform distribution, the estimate is: 
\begin{equation*}
\hat{x} = \frac{\sum_x e^{-\lambda \cdot regret_1(x,\tilde{a})} \cdot x}{\sum_x e^{-\lambda \cdot regret_1(x,\tilde{a})}}
\end{equation*}
This estimator is the one that minimizes the expected squared error for the basic quantal regret modeling assumption. 
The constant $\lambda$ is the ``regret aversion'' parameter; as $\lambda$ grows larger, the quantal regret estimate approaches the min-regret estimate suggested by \cite{eva}.

\bibliographystyle{aaai}
\bibliography{behaveconometrics}

\end{document}